\documentclass[pdflatex,sn-mathphys-num]{sn-jnl}

\usepackage{graphicx}%
\usepackage{multirow}%
\usepackage{amsmath,amssymb,amsfonts}%
\usepackage{amsthm}%
\usepackage{mathrsfs}%
\usepackage[title]{appendix}%
\usepackage{xcolor}%
\usepackage{textcomp}%
\usepackage{manyfoot}%
\usepackage{booktabs}%
\usepackage{algorithm}%
\usepackage{algorithmicx}%
\usepackage{algpseudocode}%
\usepackage{listings}%

\theoremstyle{thmstyleone}%
\theoremstyle{thmstyletwo}%

\theoremstyle{thmstylethree}%

\begin{document}

\title[Article Title]{The joint probability distribution function of global interaction complex systems}


\author{\sur{Jiaqi Zheng}}

\author{\sur{Zhifu Huang}}\email{zfhuang@hqu.edu.cn}

\affil{\orgdiv{College of Information Science and Engineering}, \orgname{Huaqiao University}, \city{Xiamen}, \country{People' Republic of China}}


\abstract{Based on the relationship that the interaction energy between any two subsystems is equal to their internal energy multiplied by the interaction coefficient, we have derived a series correlated expressions of statistical physical quantities, such as the interactional average energy, the probability distribution function of internal energy, and so on. It should be noted that the probability distribution function we obtained is existing in many complex systems involving long-range interactions. Further, based on the zeroth law of thermodynamics, we have also obtained the joint probability distribution function of two coupled complex systems. It must be note that the joint probability distribution function we obtained no longer satisfies the common independent multiplication relationship in the previous researches. Finally, we selected several representative foreign exchange high-frequency trading data for verification and found that the theory is consistent with the reality.}




\maketitle

In recent years, the research field of statistical physics has become increasingly broad. Statistical physics is the foundation of many interdisciplinary fields related to complex systems. However, the dynamics in many fields are not clear, the interaction relationships are unknown, and some cannot be obtained. Therefore, the statistical physics theory of complex systems without specific interaction forms should also be researched \cite{pres02}. In addition, for many systems with long-range interactions, previous researches have been unable to derive statistical physical properties from the basic interaction relationships. Based on this, we propose a new form of global interaction, which can derive statistical physical properties from the interactions.

Suppose the internal energy of an isolated system is $E$, which is composed of two subsystems. The internal energy of the two subsystems can be expressed as $u_{1}$ and $u_{2}$, which may be changed over time due to energy exchange. If the interaction between two subsystems can be ignored, then $E$ is a simple addition of the energy of them. However, in the real world, the interactions between subsystems should not be ignored. Here we consider the simplest case, where the interaction energy depends only on the energy of the subsystems. Inspired by the Ising model, we assume that interaction energy of two arbitrary subsystems is proportional to the product of their respective energy, where the proportional coefficient is $\lambda$. It is worth noting note that the assumption do not include many factors, such as distance or momentum. In addition, it is important to note that the interaction we present here is global.
Therefore, the internal energy $E$ can be written as
\begin{equation}
    E=u_{1}+u_{2}+\lambda u_{1} u_{2}.
    \label{eq:uu}
\end{equation}
After that, we introduced a quantity $\tilde{u}$, it means that when the internal energy of the whole system remains unchanged, the energy of each subsystem takes the same value $\tilde{u}$. Consequently, the internal energy $E$ can also be written as $E=\tilde{u}+\tilde{u}+\lambda \tilde{u} \tilde{u}.$ Using this equation, we can obtain that $\tilde{u}=(\sqrt{1+\lambda E}-1) / \lambda$. It must be mentioned that the energy $u_{1}(t)$ and $u_{2}(t)$ of each subsystem evolve over time, and their arithmetic average $\overline{u(t)}=\left[u_{1}(t)+u_{2}(t)\right] / 2$ also evolves over time. However, the quantity $\tilde{u}$ is a conserved quantity, only related to the total internal energy and interaction coefficient, and does not change with time. Thus, in order to distinguish quantity $\tilde{u}$ from arithmetic average energy, we can call this quantity as interactional average energy.

If the system has three subsystems, the total energy can be written in two ways. In case one, their total energy can be calculated as single subsystem's energy plus two subsystems' interaction energy. In case two, regarding the subsystem one and subsystem two as a whole. Comparing the above two cases, one can find that the three subsystems' interaction energy should be considered. Therefore, we have to introduce a three subsystems' interaction coefficient. In general, all possible interactions between multiple subsystems must be considered, and the m subsystems' interaction coefficient can be derived. As a result, the total internal energy of the system with N subsystems should be written. Taking logarithms, $\tilde{u}$ can be given as

\begin{equation}
    \tilde{u}=\frac{\exp \left[\sum_{u} p(u) \ln \left(1+\lambda u\right)\right]-1}{\lambda}.
    \label{eq:iu}
\end{equation}
where $p(u)$ is the probability distribution function of a subsystem with energy $u$. It must be mentioned here that, when the total internal energy of the system remains constant, the interactional average energy also remains constant. In addition, we can take the form of Shannon entropy \cite{sha1948} $S=-k_{B} \sum_{x} p(x) \ln p(x)$, which is universally applicable in various scientific fields. For the sake of convenience, we set the Boltzmann constant $k_{B}$ to unity. Considering the energy constraint and probability normalization constraint, the Lagrangian undetermined factor relation can be written as $\delta\left(S-\alpha \sum_{u} p(u)-\beta \tilde{u}\right)=0$. Thus, one can obtain probability distribution function as
\begin{equation}
    p(u)=e^{-1-\alpha}\left(1+\lambda u\right)^{-\gamma},
    \label{eq:pp}
\end{equation}
where $\gamma=\beta \exp \left(\sum_{u} p(u) \ln \left(1+\lambda u\right)\right) / \lambda$. It is very significant to note that equation ~(\ref{eq:pp}) can be expressed as q-exponential distribution which obtained in \cite{tsa1988}-\cite{tsa2009}. Press$\acute{e}$ et al. \cite{pre2013} concluded that for modeling nonexponential distributions such as power laws \cite{mfsh00}-\cite{cai01}, nonextensivity should be expressed through the constraints. By introducing the global interaction, we obtain the corresponding energy constraint relation, and derive the corresponding distribution function in accordance with Press$\acute{e}$'s view. In addition, As $u$ is large enough, equation ~(\ref{eq:pp}) can be written as well-known power-law distribution.
\begin{figure}
\includegraphics[width=12truecm]{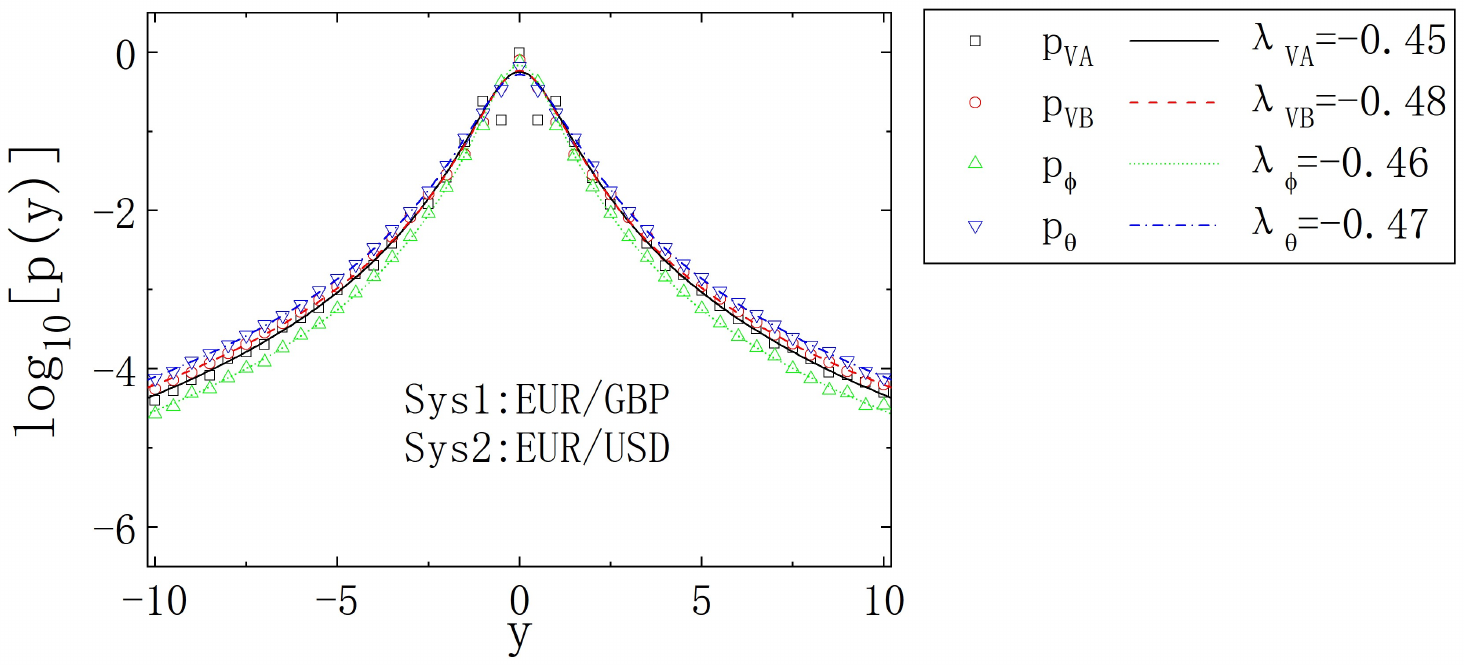}
\caption{\label{fig:nu}
The PDFs of variables.}
\end{figure}

Through the definition and derivation of the energy relationship of global interaction, the common generalized power-law distribution function form~(\ref{eq:pp}) has been obtained. Due to the fact that the product of entropy and temperature satisfies the equal relationship of the internal energy, combining with the zeroth law of thermodynamics, and the coupled global interaction energy of complex system~(\ref{eq:uu}), under the condition of equal temperature, the relationship of entropy in the coupled global interaction systems satisfies the following structure
\begin{equation}
    S_{A,B} = S_A + S_B + g S_A S_B.
    \label{eq:sg}
\end{equation}
Considering that entropy still satisfies the Shannon form, we can obtain the result that the probability distribution function of the coupled global interaction complex system satisfies the following relationship $ \sum_{x,y} p_{A,B}(x,y)[1 - g \ln p_{A,B}(x,y)] = \sum_{x} p_{A}(x)[1 - g \ln p_{A}(x)] \sum_{y} p_{B}(y)[1 - g \ln p_{B}(y)].$ Here we focus on the deviation of the principle of independent multiplication of joint probabilities, and we can define the deviation coefficient $r_{A,B}(x,y) = \frac{p_{A,B}(x,y)}{p_A(x)p_B(y)}. $ Therefore, the deviation coefficient can be derived as follows
\begin{equation}
r_{A, B}(x, y)=1+\frac{\mathrm{g}^{2} \ln p_{A}(x) \ln p_{B}(y)+\mathrm{g} \ln r_{A, B}(x, y)}{1-\mathrm{g} \ln r_{A, B}(x, y)-g \ln p_{A}(x)-g \ln p_{B}(y)}.
    \label{eq:pc}
\end{equation}

Here one can easily see that when g=0, the deviation coefficient returns to 1, which is the common principle of independent multiplication of joint probabilities. If g is not equal to 0, then the joint probability no longer satisfies the principle of independent multiplication. Next, we conduct an empirical test on the deviation coefficient of the joint probability. We select two different financial price systems and consider the relationship satisfied by the joint probability distribution function between them. We choose the minute data of the financial price system, calculate the difference of adjacent logarithmic prices per unit time, and scale-free the results $v(t)=\ln [\operatorname{price}(t+\Delta t)-\operatorname{price}(t)] /(\sigma \Delta \mathrm{t})$ \cite{mrsh00}-\cite{aab01}. Notice that, we will refer to this value as velocity.
Because there is covariance in the velocity of different systems, and the previous deviation coefficient cannot reflect the asymmetry caused by covariance, we defined two new variables, namely the average and difference of the velocity from two different systems as $\psi(t)=\frac{v_{A}(t)+v_{B}(t)}{2}$ and $\theta(t)=v_{B}(t)-v_{A}(t)$. It is noted that the two groups of variables here represent the same event, so there is an equal relationship in the joint probability
\begin{equation}
p_{v_{A}, v_{B}}(x, y)=p_{\psi, \theta}\left(\frac{x+y}{2}, y-x\right).
    \label{eq:jp1}
\end{equation}

It is noted that the covariance of the average and difference velocity of the two systems is equal to 0, as $\langle\psi(t) \theta(t)\rangle=\frac{1}{2}\left[\left\langle v_{B}^{2}(t)\right\rangle-\left\langle v_{A}^{2}(t)\right\rangle\right]=0$. Therefore, it can be considered that the corresponding conditional probability is an even function, thus there is no need to consider the asymmetry of the conditional probability. We can analyze the deviation coefficient of the joint probability of the average and difference of the velocity through real data
\begin{equation}
r_{\psi, \theta}(x, y)=1+\frac{g_{1} g_{2} \ln p_{\psi}(x) \ln p_{\theta}(y)+g_{3} \ln r_{\psi, \theta}(x, y)}{1-g_{3} \ln r_{\psi, \theta}(x, y)-g_{1} \ln p_{\psi}(x)-g_{2} \ln p_{ \theta}(y)}.
    \label{eq:jq2}
\end{equation}

\begin{figure}
\includegraphics[width=12truecm]{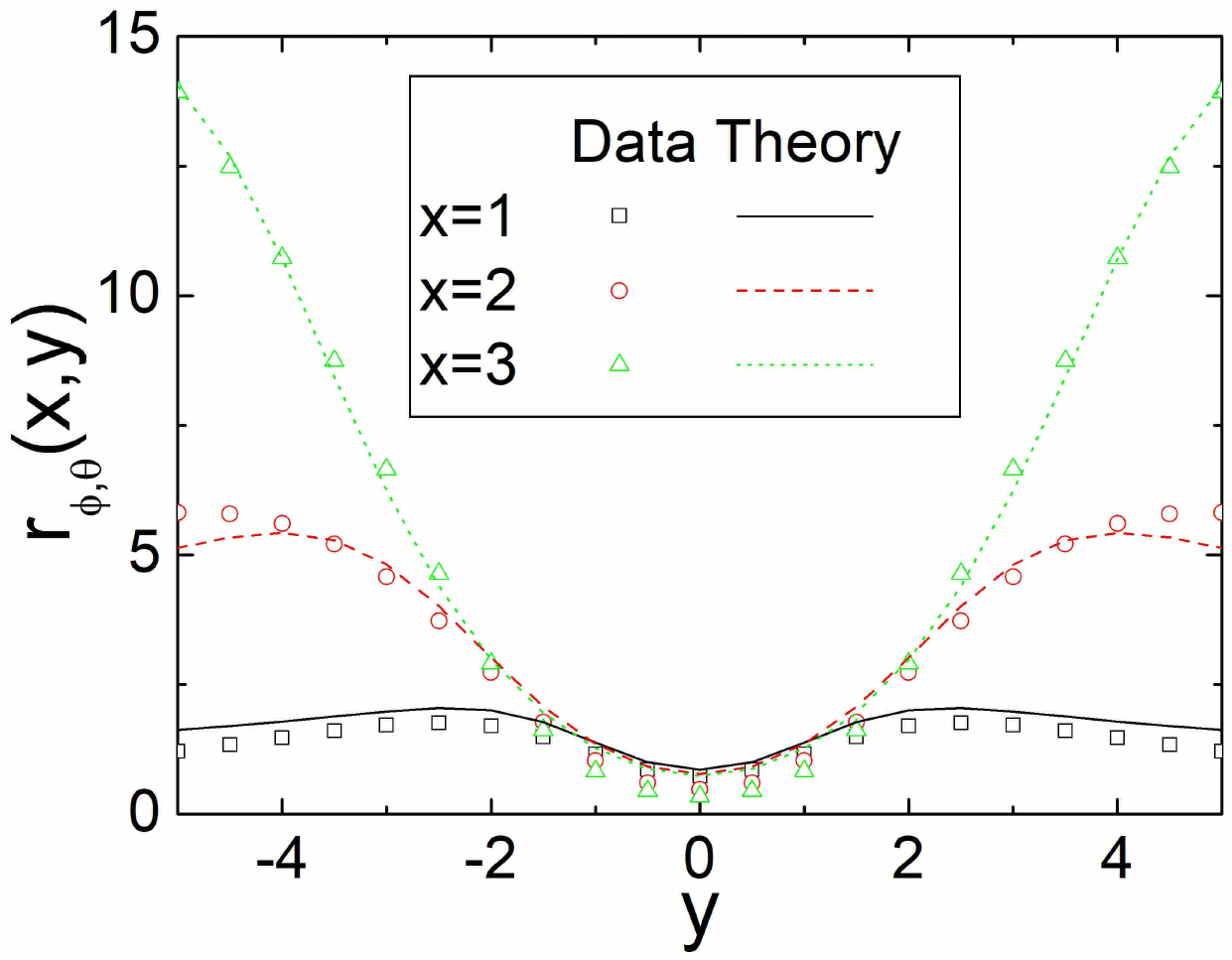}
\caption{\label{fig:jq2}
The deviation coefficient of variables $\theta$ and $\psi$.}
\end{figure}

In addition, we also know that there is a corresponding relationship between joint probability and conditional probability as follows $p_{v_{A}, v_{B}}(x, y)=p_{v_{A}}(x) p_{v_{B} \mid v_{A}}(y \mid x)$ and $p_{\psi, \theta}(x, y)=p_{\psi}(x) p_{\theta \mid \psi}(y \mid x)$. Finally, combining with equations~(\ref{eq:jp1}) and~(\ref{eq:jq2}) , we can obtain the joint probability distribution function of the velocity of two coupled systems
\begin{equation}
p_{v_{B} \mid v_{A}}(y \mid x)=\frac{r_{\psi, \theta}\left(\frac{x+y}{2}, y-x\right) p_{\psi}\left(\frac{x+y}{2}\right) p_{\theta}(y-x)}{p_{v_{A}}(x)}.
    \label{eq:jp2}
\end{equation}

\begin{figure}
\includegraphics[width=12truecm]{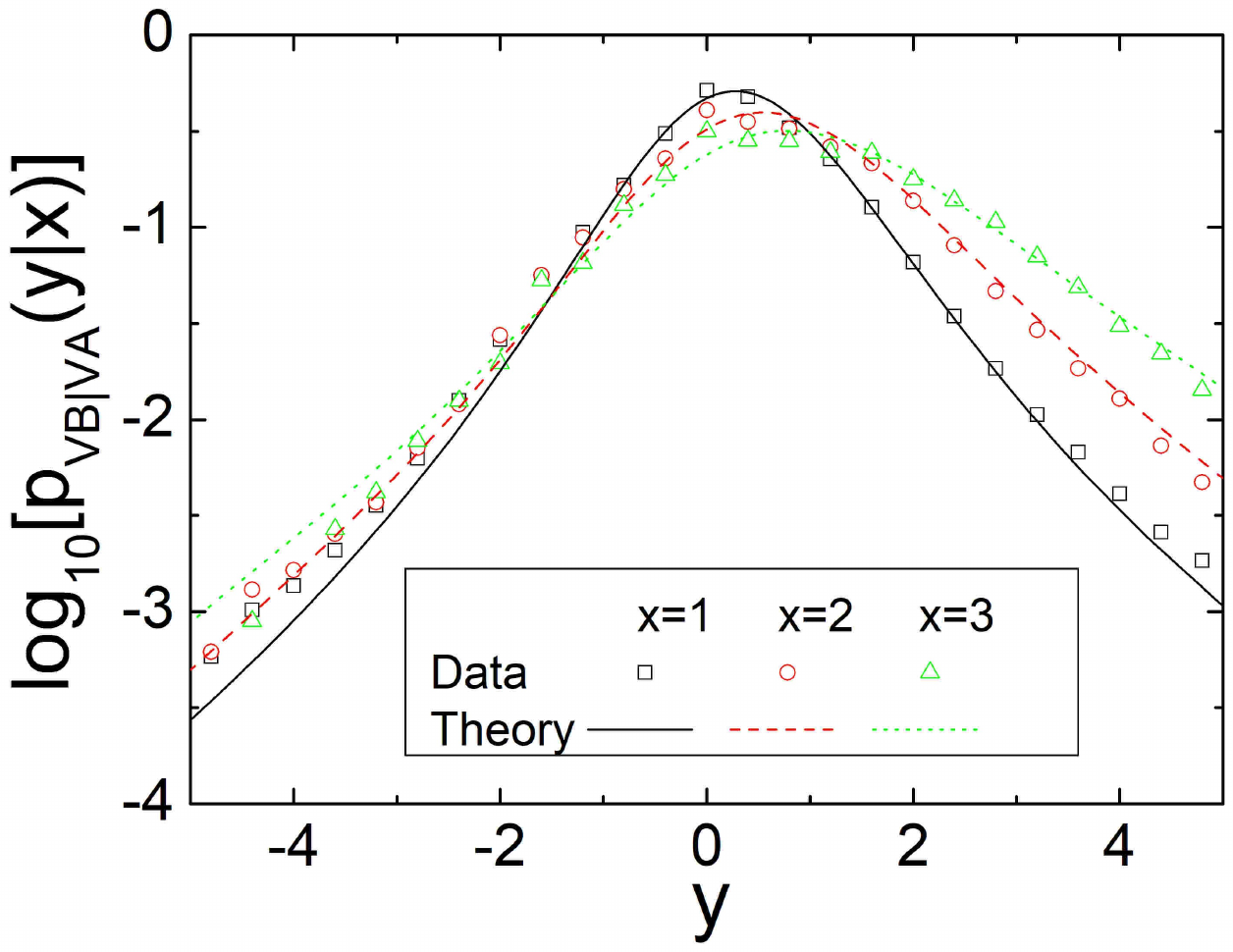}
\caption{\label{fig:jq3}
The CPDF of variables $V_{B}$ between $V_{A}$.}
\end{figure}

Based on the simple and fundamental assumption that the interaction energy of any two systems is equal to the product of their respective internal energies, we have derived several statistical physical properties of globally interacting complex systems. The distribution functions we have obtained are consistent with the power-law distribution functions that are widely present in a large number of systems. We proposed a deviation coefficient to describe the situation where the joint probability does not satisfy the principle of independent multiplication. Based on the zeroth law of thermodynamics and Shannon entropy, we obtained the expression of the deviation coefficient for the complex system with global interaction. The deviation coefficient we obtained is well agreement with the actual data. In addition, the joint probability distribution function obtained on this basis is also in good agreement with the real data. This indicates that the principle of independent multiplication of joint probability, which was widely adopted in previous researches, is not in line with reality. The joint probability distribution function we obtained may be a universal relationship. Therefore, the solution we proposed for coupling complex systems is a brand-new idea. In past research, the mainstream approach was to approximately describe interacting systems based on the idea of the mean field. In the future, through renormalization methods, it may be possible to have brand-new ideas for solving the statistical physical properties of complex systems with long-range interactions.

\bf{Contributions}

Z. H. and J. Z. designed the scheme. Z. H.  and J. Z. prepared the display items (figures). Z. H. and J. Z. co-wrote the manuscript.


\bf{Competing financial interests}

The authors declare no competing financial interests.

\renewcommand\bibname{References}

\end{document}